\documentclass[twocolumn,amsmath,amssymb,pre]{revtex4}

\usepackage{graphicx}
\usepackage{bm}

\usepackage{amsfonts}

\usepackage[utf8]{inputenc}

\begin{document}

\title{Interaction of upper hybrid waves with dust-ion-magnetoacoustic waves
and stable two-dimensional solitons in dusty plasmas}

\author{Volodymyr M. Lashkin}
\email{vlashkin62@gmail.com} \affiliation{$^1$Institute for
Nuclear Research, Pr. Nauki 47, Kyiv 03028, Ukraine}
\affiliation{$^2$Space Research Institute, Pr. Glushkova 40 k.4/1,
Kyiv 03187,  Ukraine}

\author{Oleg K. Cheremnykh}
\affiliation{Space Research Institute, Pr. Glushkova 40 k.4/1,
Kyiv 03187, Ukraine}

\begin{abstract}
We obtain a two-dimensional nonlinear system of equations for the
electrostatic potential envelope and the low-frequency magnetic
field perturbation to describe the interaction of the upper hybrid
wave propagating perpendicular to an external magnetic field with
the dust-ion-magnetoacoustic (DIMA) wave in a magnetized dusty
plasma. The equations contain both scalar and vector
nonlinearities. A nonlinear dispersion relation is derived and the
decay and modulation instability thresholds and growth rates are
obtained. Numerical estimates show that instability thresholds can
easily be exceeded in real dusty plasmas. In the static (subsonic)
approximation, a two-dimensional (2D) soliton solution (ground
state) is found numerically by the generalized Petviashvili
relaxation method. The perturbations of the magnetic field and
plasma density in the soliton are nonmonotonic in space and, along
with the perturbation in the form of a well, there are also
perturbation humps. Such peculiar radial soliton profiles differ
significantly from previously known results on 2D solitons. The
key point is that the presence of a gap in the DIMA wave
dispersion due to the Rao cutoff frequency causes the nonlinearity
to be nonlocal. We show that due to nonlocal nonlinearity the
Hamiltonian is bounded below at fixed energy, proving the
stability of the ground state.
\end{abstract}

\maketitle

\section{Introduction}

Dusty plasmas have been the subject of intensive experimental and
theoretical study for more than three decades due to their very
wide occurrence in nature
\cite{Verheest-book2000,Shukla-Mamun-book2002,Fortov2005,Melzer2021}.
They occur naturally in interstellar and interplanetary space
\cite{Gail1975,Rawlings1989}, interstellar clouds
\cite{Hartquist1997,Tsytovich2014}, planetary rings and comet
tails \cite{Angelis1988,Goertz1989,Horanyi2004}, in the Earth's
mesosphere and ionosphere \cite{Cho1993,Bernhardt1995}, etc. In
laboratory conditions, charged dust particles are present as a
contaminant in magnetic plasma confinement devices such as
tokamaks and stellarators \cite{Tsytovich1998,Krasheninnikov2008},
which negatively affects confinement. Of particular interest to
industry is dusty plasma in plasma processing environments
important for the manufacture of semiconductor devices
\cite{Boufendi2002,Bapat2007}. Another important physical
application of dusty plasma arose in connection with the
observation of Coulomb crystals in laboratory devices
\cite{Chu1994,Thomas1995,Arumugam2021}.

The properties of dusty plasma differ in many ways from usual
electron-ion plasma. Due to the fact that the mass of dust
particles exceeds the mass of ions by many orders of magnitude,
new ultra-low frequency branches of oscillations arise in the
dusty plasma, among which, for example, are the dust-acoustic wave
(DAW) theoretically predicted in Ref.~\cite{Rao1990} and
experimentally discovered in Refs.~\cite{Barkan1995,Pieper1996},
the dust-ion wave (DIW) \cite{Shukla1992,Barkan1996}, dust lattice
waves \cite{Melands1996,Homann1997}, etc. In magnetized dusty
plasmas, new modes also arise \cite{Rao1993a,Rao1993b} and, in
addition, a new characteristic plasma frequency appears, known as
the Rao cutoff frequency \cite{Rao1995}, which has no analogue in
pure electron-ion plasma. Moreover, due to the dust charging
effect, both linear modes and nonlinear structures can be
drastically modified
\cite{Nejoh1997,Wan2006,Xie1998,Tribeche2002}.

The linear theory of waves in nonmagnetized and magnetized dusty
plasmas has been developed in sufficient detail. Nonlinear
structures such as solitons, shocks and rogue waves have also been
studied in a fairly large number of both theoretical (see, e.g.
Refs.~\cite{Verheest-book2000,Shukla-Mamun-book2002,Shukla2003,Shukla2009}
and references therein, and also Ref.~\cite{Rahman2018}) and
experimental \cite{Prasad2008expPRL,Merlino2012exp,
Liu2018exp,Bandyopadhyay2022exp} works. The overwhelming majority
of studies, however, dealt with one-dimensional (1D) structures in
dusty plasmas. Multidimensional nonlinear structures in dusty
plasmas have been studied to a much lesser extent. For dusty
plasma, the two-dimensional (2D) Kadomtsev-Petviashvili (KP)
equation was obtained by the reductive perturbation method in
Refs.~\cite{Duan-KP2002,Labany-KP2004,Saini-KP2015}, but in these
cases, by replacing variables, this equation was actually reduced
to the Korteweg-de Vries equation and the corresponding solution
depends only on one effective variable, although in fact there are
also truly 2D solutions of the KP equation (the so-called lumps).
Dust solitons within the framework of cylindrical and spherical KP
equations were considered in Refs.~\cite{Moslem2010,Gaoa2006}. For
dusty plasma, the Davey-Stewartson equations
\cite{Annou2012,Saini2016} and the Zakharov-Kuznetsov equation
\cite{Song-ZK-2D-2020} were also derived. In
Refs.~\cite{Annou2012,Saini2016}, analytical true 2D solutions in
the form of so-called dromions were presented, and in
Ref.~\cite{Song-ZK-2D-2020} the dust 2D soliton was found
numerically. Two-dimensional dust dipole and tripole vortices and
vortex chains were found analytically in
Refs.~\cite{Bharuthram1992,Vranjes1999,Stenflo2003} using a
technique similar to the Larichev-Reznik method for the
atmospheres of rotating planets and magnetized electron-ion
plasmas \cite{Petviashvili_book1992} (for experimental works on
vortices in dusty plasmas see a recent review \cite{Choudhary2024}
and references therein). The lower intensity of study on
multidimensional nonlinear structures, compared to 1D ones, can be
partly explained by the fact that, as is known, such structures
often (but not always, see the examples mentioned above) turn out
to be unstable and lead either to collapse or wave breaking.

In this paper, we derive a 2D nonlinear system of equations for
the electrostatic potential envelope and the low-frequency
magnetic field perturbation to describe the interaction of the
upper hybrid (UH) wave propagating perpendicular to an external
magnetic field with the dust-ion-magnetoacoustic (DIMA) wave in a
magnetized dusty plasma. A similar 1D problem of the interaction
of an upper hybrid wave with a modified Alfv\'{e}n wave in a dusty
plasma was studied in Ref.~\cite{Shukla2003POP}. The DIMA wave was
theoretically predicted by Rao in Ref.~\cite{Rao1995}. A
distinctive feature of the DIMA wave, in contrast to other
acoustic modes in both purely electron-ion and dusty plasmas, is
the presence of a gap in its dispersion (the so-called Rao cutoff
frequency), that is, the frequency of the wave at zero wave vector
is not equal to zero and is equal to the cutoff frequency. To
avoid misunderstandings, we note that the term DIMA introduced in
Ref.~\cite{Rao1995} corresponds precisely to a dust
magnetoacoustic wave with a gap in the dispersion, in contrast to
the previously introduced in Refs.~\cite{Rao1993a,Rao1993b} term
dust magnetoacoustic wave (DMA), where there is no gap in the
dispersion (and, accordingly, there is no cutoff frequency). The
term DIMA is related to the fact that in this case only ions play
an active role in the dynamics, while dust particles are
considered to be immobile. We show that the presence of a gap in
the DIMA dispersion due to the Rao cutoff frequency results in the
nonlinearity in the resulting equations being essentially
nonlocal, i.e. the nonlinear response depends on the wave
intensity in some spatial region. An important general property of
nonlinearity with nonlocal response is that in many cases it
prevents the catastrophic collapse of multidimensional wave
packets that typically occurs in local self-focusing media with
cubic nonlinearity. In particular, a rigorous proof of the absence
of collapse in a nonlocal nonlinear Schr\"{o}dinger (NLS) equation
model with a sufficiently general symmetric real response kernel
was presented in Refs.~\cite{Turitsyn1985, Krolikovski2004}.
Moreover, it was shown that nonlocal nonlinearity arrests the
collapse and results in the existence of stable coherent
structures that collapse in models with cubic local response. A
variety of physical models were considered, including optical
media \cite{Lashkin2006,Torner2006}, plasmas with thermal
nonlinearity \cite{Yakim2005,Lashkin2007PLA}, quantum plasmas
\cite{Sulem2009}, Bose-Einstein condensates with dipole nonlocal
nonlinearity \cite{Lashkin2007PRA,Lashkin2009PhysScr}, etc. In the
presented paper we numerically find solutions in the form of a
dusty nonlocal 2D soliton (ground state). The magnetic field and
plasma density perturbations have the shape of a well with two
humps. We show that due to nonlocal nonlinearity the Hamiltonian
is bounded below at fixed energy, thus proving the stability of
the ground state.

The paper is organized as follows. In Sec.~II, we derive a system
of nonlinear equations to describe the interaction of the UH wave
with the DIMA wave. The instability of a plane wave within the
framework of the obtained equations is studied in Sec.~III.
Numerical solutions in the form of 2D solitons are found in
Sec.~IV. In Sec.~V, the stability of the 2D solitons is proved.
Finally, Sec.~VI concludes the paper.

\section{\label{model} Derivation of model equations}

We consider a homogeneous dusty plasma in a uniform external
magnetic field $\mathbf{B}_{0}=B_{0}\hat{\mathbf{z}}$, where
$\hat{\mathbf{z}}$ is the unit vector along the $z$-direction. The
linear dispersion relation for UH waves propagating almost
perpendicular to the external magnetic field, that is provided
\begin{equation}
\label{cond-perp} \frac{k_{z}}{k_{\perp}}\ll \frac{m_{e}}{m_{i}},
\end{equation}
where $k_{z}$ and $k_{\perp}$ are the wave number along the
external magnetic field and perpendicular wave number,
respectively, $m_{e}$ is the electron mass, and $m_{i}$ is the ion
mass, is
\begin{equation}
\label{disp-UH}
\omega=\omega_{uh}\left(1+\frac{1}{2}k_{\bot}^{2}R^{2}\right),
\end{equation}
where $\omega_{uh}=(\omega_{pe}^{2}+\Omega_{e}^{2})^{1/2}$ is the
UH resonance frequency, $\omega_{pe}$ is the electron plasma
frequency, $\Omega_{e}$ is the electron gyrofrequency, $v_{Te}$ is
the electron thermal speed, and
$R^{2}=3v_{Te}^{2}/\omega_{uh}^{2}$. Note that the consistent
kinetic treatment leads to an additional factor in the dispersion
term (modifying the dispersion length $R$) if the electron plasma
frequency is close enough to the electron gyrofrequency, but in
this paper we restrict ourselves to dispersion Eq.
(\ref{disp-UH}), which is justified in most real physical
situations. The upper hybrid wave is a high frequency (HF) wave
and only electrons take part in the motion.

Nonlinear equations to describe the interaction of UH waves with
low-frequency (LF) density and magnetic field perturbations in the
three-dimensional (3D) case were obtained in
Ref.~\cite{Lashkin2007}. In the 2D case, corresponding to the
almost perpendicular propagation of UH waves, and valid under the
condition Eq. (\ref{cond-perp}), equation for the slow varying
complex amplitude $\varphi$ of the potential of the HF
electrostatic electric field
\begin{equation}
\label{EH}
\mathbf{E}^{H}=-\frac{1}{2}[\nabla\varphi\exp(-i\omega_{uh}t)+\mathrm{c.c.}],
\end{equation}
where $\mathrm{c.c.}$ stands for the complex conjugation, has the
form
\begin{gather}
\label{main}
\Delta\left(\frac{2i}{\omega_{uh}}\frac{\partial\varphi}{\partial
t }+R^{2}\Delta\varphi\right) \\ \nonumber
 =\frac{1}{\omega_{uh}^{2}}\nabla\cdot\left\{\left(\omega_{pe}^{2}\frac{\tilde{n}_{e}}{n_{0e}}+
 2\Omega_{e}^{2}\frac{\tilde{B}}{B_{0}}\right)\nabla\varphi \right.
 \\ \nonumber
\left.
-i\frac{\Omega_{e}}{\omega_{uh}}\left[\omega_{pe}^{2}\frac{\tilde{n}_{e}}{n_{0e}}
+(\omega_{pe}^{2}+2\Omega_{e}^{2})\frac{\tilde{B}}{B_{0}}\right]
(\nabla\varphi\times\hat{\mathbf{z}})\right\},
\end{gather}
where $\tilde{B}$ and $\tilde{n}_{e}$ are the magnetic field and
electron plasma density perturbations, respectively. Here and
throughout the paper, $\nabla=(\partial/\partial
x,\partial/\partial y)$, $\Delta=\partial^{2}/\partial
x^{2}+\partial^{2}/\partial y^{2}$ is the 2D Laplacian, and
accordingly the subscript $\perp$ is omitted in wave numbers and
vectors. When deriving Eq. (\ref{main}) in
Ref.~\cite{Lashkin2007}, the physical meaning of LF perturbations
of the plasma density $\delta n$ and magnetic field $\delta B$ was
generally not specified (specifically, in Ref.~\cite{Lashkin2007},
the kinetic Alfv\'{e}n wave was then considered as a low-frequency
perturbation). In fact, this equation was obtained using the
substitutions
\begin{equation}
\omega_{pe}\rightarrow
\omega_{pe}\left(1+\frac{\tilde{n}_{e}}{n_{0e}}\right), \quad
\mathrm{and}  \quad
\Omega_{e}\rightarrow\Omega_{e}\left(1+\frac{\tilde{B}}{B_{0}}\right)
\end{equation}
in the linear dielectric tensor of magnetized plasma (taking into
account a weak thermal dispersion), where $\tilde{n}_{e}$ and
$\tilde{B}$ account for the corresponding nonlinear frequency
shifts.

For the LF dust-ion-magnetoacoustic mode, dust particles are
assumed to be almost immobile \cite{Rao1995}, and then we proceed
from the equations of motion for ions,
\begin{equation}
\label{ion-eq} m_{i}\frac{\partial \mathbf{v}_{i}}{\partial
t}=e\mathbf{E}+\frac{eB_{0}}{c}(\mathbf{v}_{i}\times\hat{\mathbf{z}})-\frac{\gamma_{i}T_{i}\nabla
n_{i}}{n_{i0}}
\end{equation}
and inertialess electrons,
\begin{equation}
\label{inertionless}
\mathbf{F}=-e\mathbf{E}-\frac{eB_{0}}{c}(\mathbf{v}_{e}\times\hat{\mathbf{z}})
-\frac{\gamma_{e}T_{e}\nabla n_{e}}{n_{e0}},
\end{equation}
where
\begin{equation}
\label{F-pon} \mathbf{F}=m_{e}\langle
(\mathbf{v}^{H}_{e}\cdot\nabla)\mathbf{v}^{H}_{e}\rangle
+\left\langle\frac{e}{c}[\mathbf{v}^{H}_{e}\times\mathbf{B}^{H}]\right\rangle
\end{equation}
is the ponderomotive force acting on electrons, and
$\mathbf{v}_{e}^{H}$ and $\mathbf{B}^{H}$ are the HF electron
velocity and magnetic field perturbation, respectively,
\begin{gather}
\label{vH}
\mathbf{v}_{e}^{H}=\frac{1}{2}[\tilde{\mathbf{v}}\exp(-i\omega_{uh}t)+\mathrm{c.c.}],
\\
\label{BH}
\mathbf{B}^{H}=\frac{1}{2}[\tilde{\mathbf{B}}\exp(-i\omega_{uh}t)+\mathrm{c.c.}].
\end{gather}
The angular brackets in Eq. (\ref{F-pon}) denote averaging over
the HF oscillations. In Eqs. (\ref{ion-eq}) and
(\ref{inertionless}) the notations for particles of species
$\alpha$ are used ($\alpha=e,i$ - electrons and ions), so that
$\mathbf{v}_{\alpha}$ is the particle velocity,
$n_{\alpha}=n_{\alpha 0}+\tilde{n}_{\alpha}$ is the particle
density, $n_{\alpha 0}$ and $\tilde{n}_{\alpha}$ are the
corresponding equilibrium and perturbed particle densities,
$T_{\alpha}$ is the temperature and $\gamma_{\alpha}$ is the ratio
of specific heats. In Eqs. (\ref{vH}) and (\ref{BH}),
$\tilde{\mathbf{v}}$ and $\tilde{\mathbf{B}}$ are the envelopes of
the corresponding quantities at the UH frequency $\omega_{uh}$.
Equations (\ref{ion-eq}) and (\ref{inertionless}) are closed by
the continuity equations for ions and electrons,
\begin{gather}
\frac{\partial n_{i}}{\partial t}+n_{i0}\nabla\cdot
\mathbf{v}_{i}=0,
 \label{ion-contin}
 \\
 \frac{\partial n_{e}}{\partial t}+n_{e0}\nabla\cdot
\mathbf{v}_{e}=0,
 \label{electron-contin}
\end{gather}
and the Maxwell equations,
\begin{equation}
\label{Maxwell11} \hat{\mathbf{z}}\cdot
(\nabla\times\mathbf{E})=-\frac{1}{c}\frac{\partial
\tilde{B}}{\partial t},
\end{equation}
\begin{equation}
\label{Maxwell2} \nabla\times\mathbf{B}=\frac{4\pi
e}{c}(n_{i0}\mathbf{v}_{i}-n_{e0}\mathbf{v}_{e}),
\end{equation}
where we have neglected the displacement current for the LF
motion. As noted below, dust particles do not participate in the
motion but provide overall charge neutrality of the plasma,
\begin{equation}
\label{neutral} n_{i0}=n_{e0}+Z_{d}n_{d0},
\end{equation}
where $Z_{d}$ is the number (taking into account the sign) of the
charge residing on the dust grains ($Z_{d}>0$ for positively
charged dust particles and $Z_{d}<0$ for negatively charged ones),
and $n_{d0}$ is the equilibrium dust density. The only nonlinear
effect is the presence of a ponderomotive force in Eq.
(\ref{inertionless}). To calculate the ponderomotive force in Eq.
(\ref{F-pon}), we use the HF equation of motion for electrons,
\begin{equation}
\label{HF-electrons} m_{e}\frac{\partial
\mathbf{v}_{e}^{H}}{\partial
t}=-e\mathbf{E}^{H}-\Omega_{e}[\mathbf{v}_{e}^{H}\times\hat{\mathbf{z}}]
\end{equation}
and then, substituting Eqs. (\ref{EH}) and (\ref{vH}) into Eq.
(\ref{HF-electrons}), one can obtain
\begin{equation}
\label{v2}
\tilde{\mathbf{v}}=\frac{e}{m}\frac{[i\omega_{uh}\nabla\varphi+
\Omega_{e}(\nabla\varphi\times\hat{\mathbf{z}})]}{\omega_{pe}^{2}}.
\end{equation}
Taking into account that in the zero order,
\begin{equation}
\frac{\partial}{\partial
t}(\nabla\times\mathbf{v}^{H}_{e})=-\frac{e}{m_{e}}(\nabla\times\mathbf{E}^{H}),
\end{equation}
and using the Maxwell equation for $\partial
\mathbf{B}^{H}/\partial t$, we find
\begin{equation}
\label{BH2}
\mathbf{B}^{H}=\frac{m_{e}c}{e}(\nabla\times\mathbf{v}_{e}^{H}).
\end{equation}
With the aid of Eq. (\ref{BH2}), two terms in Eq. (\ref{F-pon})
for $\mathbf{F}$ can be combined to yield
\begin{gather}
\mathbf{F}=m_{e}\langle
(\mathbf{v}_{e}^{H}\cdot\nabla)\cdot\mathbf{v}_{e}^{H}+
[\mathbf{v}_{e}^{H}\times[\nabla\times\mathbf{v}_{e}^{H}]]\rangle
\nonumber \\
=\frac{m_{e}}{2}\langle\nabla\,(\mathbf{v}_{e}^{H}\cdot\mathbf{v}_{e}^{H})\rangle
 =\frac{m_{e}}{4}\nabla\,|\tilde{\mathbf{v}}|^{2}.
 \label{F}
\end{gather}
Inserting $\tilde{\mathbf{v}}$ from Eq. (\ref{v2}) into Eq.
(\ref{F}) we have
\begin{equation}
\label{F1}
\mathbf{F}=\frac{e^{2}}{4\omega_{pe}^{4}}\nabla[(\omega_{uh}^{2}+\Omega_{e}^{2})|\nabla\varphi|^{2}
+2i\omega_{uh}\Omega_{e}\{\varphi,\varphi^{\ast}\}],
\end{equation}
where we have introduced the notation for the Poisson bracket
(Jacobian)
\begin{equation}
\label{bracket} \{f,g\}=\frac{\partial f}{\partial
x}\frac{\partial g}{\partial y}-\frac{\partial f}{\partial
y}\frac{\partial g}{\partial x}\equiv [\nabla f\times \nabla
g]\cdot\hat{\mathbf{z}}.
\end{equation}
Note that the second term in Eq. (\ref{F1}) is real. Solving Eq.
(\ref{Maxwell2}) for $\mathbf{v}_{e}$ and substituting it into Eq.
(\ref{inertionless}) we obtain for $\mathbf{E}$,
\begin{equation}
\mathbf{E}=\frac{B_{0}}{cn_{e0}}\left(\frac{c}{4\pi
e}\nabla\times\mathbf{B}-n_{i0}\mathbf{v}_{i}\right)\times\hat{\mathbf{z}}
 -\frac{\gamma_{e}T_{e}}{en_{e0}}\nabla n_{e}-\frac{\mathbf{F}}{e}.
\end{equation}
Inserting $\mathbf{E}$ into equation of motion for ions Eq.
(\ref{ion-eq}) one can get
\begin{gather}
\frac{\partial \mathbf{v}_{i}}{\partial
t}=-\Omega_{R}(\mathbf{v}_{i}\times\hat{\mathbf{z}})+\frac{B_{0}}{4\pi
m_{i}n_{e0}}(\nabla\times\mathbf{B})\times\hat{\mathbf{z}}
\nonumber \\
-\frac{\gamma_{e}T_{e}\nabla n_{e}}{m_{i}n_{e0}}
-\frac{\gamma_{i}T_{i}\nabla
n_{i}}{m_{i}n_{i0}}-\frac{\mathbf{F}}{m_{i}}, \label{eq3}
\end{gather}
where we used charge neutrality Eq. (\ref{neutral}), and
\begin{equation}
\Omega_{R}=\frac{Z_{d}n_{d0}\Omega_{i}}{n_{e0}},
\end{equation}
is the Rao cutoff frequency, and $\Omega_{i}$ is the ion
gyrofrequency. Taking divergence of Eq. (\ref{eq3}) and then using
Eq. (\ref{ion-contin}) we have
\begin{gather}
\frac{\partial^{2}n_{i}}{\partial
t^{2}}-n_{i0}\Omega_{R}\hat{\mathbf{z}}\cdot
(\nabla\times\mathbf{v}_{i})-\frac{B_{0}n_{i0}}{4\pi
m_{i}n_{e0}}\Delta \tilde{B}
\nonumber \\
-\frac{\gamma_{e}T_{e}n_{i0}}{m_{i}n_{e0}}\Delta n_{i}
-\frac{\gamma_{i}T_{i}}{m_{i}}\Delta
n_{i}=\frac{n_{i0}}{m_{i}}\nabla\cdot\mathbf{F}, \label{eq4}
\end{gather}
where it has been taken into account that $\mathbf{B}=\tilde{B}
\hat{\mathbf{z}}$. Substituting $\mathbf{E}$ from Eq.
(\ref{ion-eq}) into Eq. (\ref{Maxwell11}) and eliminating
$\nabla\cdot\mathbf{v}_{i}$ with the aid of Eq. (\ref{ion-contin})
we get
\begin{equation}
\label{delta-B} \tilde{B}=B_{0}\frac{\tilde{
n}_{i}}{n_{i0}}-\frac{m_{i}c}{e}\hat{\mathbf{z}}\cdot
(\nabla\times\mathbf{v}_{i}).
\end{equation}
Taking curl of of Eq. (\ref{inertionless}) and eliminating
$\nabla\cdot\mathbf{v}_{e}$ in the resulting equation with the aid
of Eq. (\ref{electron-contin}), we substitute
$\nabla\times\mathbf{E}$ into Eq. (\ref{Maxwell11}) and find  the
frozen-in-field relation
\begin{equation}
\label{frozen}
\frac{\tilde{B}}{B_{0}}=\frac{\tilde{n}_{e}}{n_{e0}}.
\end{equation}
On the other hand, Eq. (\ref{Maxwell2}) implies
$n_{i0}\nabla\cdot\mathbf{v}_{i}=n_{e0}\nabla\cdot\mathbf{v}_{e}$,
and from Eqs. (\ref{ion-contin}) and (\ref{electron-contin}) we
immediately get $\tilde{n}_{i}=\tilde{n}_{e}$. Taking this into
account and Eq. (\ref{eq4}) together with Eqs. (\ref{delta-B}) and
(\ref{frozen}), one can get
\begin{equation}
\label{LF11} \frac{\partial^{2}\tilde{B}}{\partial
t^{2}}+\Omega_{R}^{2}\tilde{B}-(v_{A}^{2}+v_{s}^{2})\Delta
\tilde{B}=\frac{B_{0}}{m_{i}}\nabla\cdot\mathbf{F},
\end{equation}
where $v_{A}$ and $v_{s}$ are the modified Alfv\'{e}n velocity and
acoustic velocity, respectively,
\begin{equation}
v_{A}^{2}=\frac{n_{i0}B_{0}^{2}}{4\pi n_{e0}^{2}m_{i}}, \quad
v_{s}^{2}=\frac{n_{i0}\gamma_{e}T_{e}}{n_{e0}m_{i}}+\frac{\gamma_{i}T_{i}}{m_{i}}.
\end{equation}
Equation (\ref{LF11})  describe the dynamics of LF acoustic-type
disturbances (in the linear case corresponding to the DIMA wave)
under the action of the ponderomotive force of the HF field of UH
wave. Inserting Eq. (\ref{frozen}) into Eq. (\ref{main}), we
finally obtain the equation for the electrostatic potential
envelope $\varphi$,
\begin{gather}
\label{mainH}
\Delta\left(\frac{2i}{\omega_{uh}}\frac{\partial\varphi}{\partial
t }+R^{2}\Delta\varphi\right) \\ \nonumber
 =\nabla\cdot\left\{\frac{\tilde{B}}{B_{0}}\left[\left(1+\frac{\Omega_{e}^{2}}{\omega_{uh}^{2}}
 \right)\nabla\varphi
 -2i\frac{\Omega_{e}}{\omega_{uh}}
(\nabla\varphi\times\hat{\mathbf{z}})\right]\right\}.
\end{gather}
In turn, from Eqs. (\ref{F1}) and (\ref{LF11} )  we find the
equation for the LF magnetic field perturbation,
\begin{gather}
\frac{\partial^{2}\tilde{B}}{\partial
t^{2}}+\Omega_{R}^{2}\tilde{B}-(v_{A}^{2}+v_{s}^{2})\Delta
\tilde{B}
\nonumber \\
=\frac{e^{2}B_{0}}{4\omega_{pe}^{4}m_{i}}\Delta
[(\omega_{uh}^{2}+\Omega_{e}^{2})|\nabla\varphi|^{2}
+2i\omega_{uh}\Omega_{e}\{\varphi,\varphi^{\ast}\}]. \label{mainL}
\end{gather}
Equations (\ref{mainH}) and (\ref{mainL}) are a closed system of
equations to describe the interaction of HF upper hybrid waves
with LF dust-ion-magnetoacoustic waves in a dusty magnetized
plasma. In the linear approximation, Eqs. (\ref{mainH}) and
(\ref{mainL}) give the dispersion relation for the UH wave
(\ref{disp-UH}), and the dispersion relation for the DIMA wave
\begin{equation}
\omega^{2}=\Omega_{R}^{2}+k^{2}(v_{A}^{2}+v_{s}^{2}),
\end{equation}
respectively. The dispersion of the DIMA  wave is of the so-called
optical type ($\omega\rightarrow \omega_{c}$ as
$\mathbf{k}\rightarrow 0$, where $\omega_{c}$ is the cutoff
frequency), that is, there is a gap determined by the Rao cutoff
frequency $\Omega_{R}$, in contrast to the dispersion of
conventional acoustic waves ($\omega \rightarrow 0$ as
$\mathbf{k}\rightarrow 0$). It should be noted that in
Ref.~\cite{Rao1995}, without loss of generality, a 1D case of DIMA
wave propagation perpendicular to the external magnetic field was
considered with spatial dependence only on the $x$ coordinate.
After introducing the corresponding dimensionless variables,
\begin{gather}
t\rightarrow \frac{\omega_{uh}t}{2}, \quad \mathbf{r}\rightarrow
\frac{\mathbf{r}}{R},
\\
b=\frac{\tilde{B}}{B_{0}}, \quad \varphi\rightarrow
\frac{e\,\omega_{uh}}{2\omega_{pe}^{2}R\sqrt{m_{i}(v_{A}^{2}+v_{s}^{2})}}\,\varphi
,
\\
\alpha=\frac{\Omega_{R}^{2}R^{2}}{v_{A}^{2}+v_{s}^{2}}, \quad
\beta=\frac{\omega_{uh}^{2}}{4\Omega_{R}^{2}}\alpha, \quad
\mu=\frac{\Omega_{e}}{\omega_{uh}},
 \end{gather}
equations (\ref{mainH}) and (\ref{mainL}) become
\begin{equation}
\label{mainHFdim} \Delta\left(i\frac{\partial\varphi}{\partial t
}+\Delta\varphi\right)
 =\nabla\cdot\left\{b[(1+\mu^{2})\nabla\varphi-2i\mu (\nabla\varphi\times\hat{\mathbf{z}})]\right\},
 \end{equation}
and
\begin{equation}
\label{mainLFdim}
 \beta\frac{\partial^{2}b}{\partial t^{2}}+\alpha b-\Delta b=\Delta
 [(1+\mu^{2})|\nabla\varphi|^{2}+2i\mu
 \{\varphi,\varphi^{\ast}\}],
 \end{equation}
respectively. It is evident that these equations contain both
scalar and vector nonlinearities (the latter is absent in the 1D
case), which, generally speaking, can be of the same order.

\section{\label{instability}Instability of a plane wave}

To study the linear stage of instability within the framework of
nonlinear Eqs. (\ref{mainHFdim}) and (\ref{mainLFdim}) we
decompose the UH wave into the pump wave of the form of plane wave
with the amplitude $\varphi_{0}$ and wave vector $\mathbf{k}$, and
two sideband perturbations corresponding to the linear modulation
with the frequency $\Omega$ and wave vector $\mathbf{q}$,
\begin{gather}
\varphi=\varphi_{0}\mathrm{e}^{i\mathbf{k}\cdot\mathbf{r}-i\omega_{\mathbf{k}}t}
+\varphi_{+}\mathrm{e}^{i(\mathbf{k}
+\mathbf{q})\cdot\mathbf{r}-i(\omega_{\mathbf{k}}+\Omega)t}
\nonumber \\
+\varphi_{-}\mathrm{e}^{i(\mathbf{k}
-\mathbf{q})\cdot\mathbf{r}-i(\omega_{\mathbf{k}}-\Omega)t} +
\mathrm{c. c.}, \label{phi-high}
\end{gather}
where $\omega_{\mathbf{k}}=k^{2}$ is the frequency of the plane
wave, which is an exact solution of Eqs. (\ref{mainHFdim}) and
(\ref{mainLFdim}). The LF perturbations of the magnetic field is
expressed as
\begin{equation}
\label{B-low}
b=\hat{b}\mathrm{e}^{i\mathbf{q}\cdot\mathbf{r}-i\Omega t}+
\mathrm{c. c.} .
\end{equation}
By linearizing with respect to perturbations, one can readily
calculate the satellite amplitudes $\varphi_{\pm}$ using Eq.
(\ref{mainHFdim}), and we have,
\begin{gather}
\label{four1} D_{+}\varphi_{+}=A_{+}\hat{b}\varphi_{0},
\\
\label{four2}
D_{-}\varphi_{-}^{\ast}=A_{-}\hat{b}\varphi_{0}^{\ast},
\end{gather}
where the coefficients $D_{\pm}$ and $A_{\pm}$ are given by
\begin{gather}
D_{\pm}=(\mathbf{k}\pm\mathbf{q})^{2}[(\mathbf{k}\pm\mathbf{q})^{2}-k^{2}\mp\Omega]
\nonumber \\
=\omega_{\mathbf{k}\pm\mathbf{q}}[\omega_{\mathbf{k}\pm\mathbf{q}}-\omega_{\mathbf{k}}\mp\Omega],
\label{D-pm}
\\
A_{\pm}=-(1+\mu^{2})(k^{2}\pm \mathbf{k}\cdot\mathbf{q})-2i\mu
(\mathbf{k}\times\mathbf{q})_{z}.
\end{gather}
The amplitude of the LF perturbation $\hat{b}$ is found from Eq.
(\ref{mainLFdim}),
\begin{equation}
\label{four3}
(\Omega^{2}-\Omega^{2}_{q})\hat{b}=\frac{q^{2}}{\beta}(B_{+}\varphi_{+}\varphi_{0}^{\ast}
+B_{-}\varphi_{-}^{\ast}\varphi_{0}),
\end{equation}
where $\Omega_{q}$ corresponds to the linear dispersion of the LF
wave,
\begin{gather}
\label{Om-q}
\Omega^{2}_{q}=\frac{\alpha+q^{2}}{\beta},
\\
\label{Bpm}
 B_{\pm}=(1+\mu^{2})(k^{2}\pm\mathbf{k}\cdot\mathbf{q})-2i\mu(\mathbf{k}\times\mathbf{q})_{z}.
\end{gather}
Combining Eqs. (\ref{four1}), (\ref{four2}) and (\ref{four3}) we
have a nonlinear dispersion relation,
\begin{equation}
\label{non-disp}
\Omega^{2}-\Omega^{2}_{q}=\frac{q^{2}|\varphi_{0}|^{2}}{\beta}\left(\frac{C_{+}}{D_{+}}
+\frac{C_{-}}{D_{-}}\right),
\end{equation}
where
\begin{equation}
\label{C-pm}
C_{\pm}=A_{\pm}B_{\pm}=-\!\left[(1+\mu^{2})^{2}(k^{2}\pm\mathbf{k}\cdot\mathbf{q})^{2}
+4\mu^{2}(\mathbf{k}\times\mathbf{q})_{z}^{2}\right].
\end{equation}
The terms with scalar and vector products correspond to scalar and
vector nonlinearity in nonlinear Eqs. (\ref{mainHFdim}) and
(\ref{mainLFdim}), respectively. Equation (\ref{non-disp}) is a
quartic equation with real coefficients in $\Omega$ and can be
solved exactly. In the case of complex roots (conjugate pair), the
dispersion relation Eq. (\ref{non-disp}) predicts instability with
the growth rate $\gamma=|\mathrm{Im}\,\Omega|$. The stability and
instability regions (as well as the corresponding growth rate)
depend strongly not only on the pump amplitude $\varphi_{0}$, but
also on the relationship between the wave vector of the plane pump
wave $\mathbf{k}$ and the wave vector of the perturbation
$\mathbf{q}$. In the case $\mathbf{k}\parallel\mathbf{q}$, the
parametric coupling of the waves due to the vector nonlinearity is
absent, while the coupling due to the scalar nonlinearity is the
most effective. In the opposite case $\mathbf{k}\perp\mathbf{q}$,
the interaction due to the vector nonlinearity is the most
effective, while the interaction due to the scalar nonlinearity is
weakened (and almost absent if $k\ll q$). In the general case and
when $\mu\sim 1$, the overall picture turns out to be quite
complex. A detailed study of Eq. (\ref{non-disp}) is beyond the
scope of this paper and we restrict ourselves to a number of
special cases.

\subsection{Decay instability}
At not too large pump amplitudes, the second term in brackets in
Eq. (\ref{non-disp}) is resonant ($\omega_{\mathbf{k}}\sim
\omega_{\mathbf{k}-\mathbf{q}}+\Omega$) and is significantly
larger than the first one. This corresponds to the excitation of
only one of the satellites and the so-called decay instability.
The nonlinear dispersion relation Eq. (\ref{non-disp}) then
reduces to
\begin{gather}
(\Omega^{2}-\Omega^{2}_{q})[k^{2}-(\mathbf{k}-\mathbf{q})^{2}-\Omega]=\frac{q^{2}|E_{0}|^{2}}{\beta}
\nonumber \\
\times[(1+\mu^{2})^{2}\cos^{2}\theta+4\mu^{2}\sin^{2}\theta],
\label{non-disp-decay1}
\end{gather}
where $E_{0}$ is the electric pump field,
$|E_{0}|^{2}=k^{2}|\varphi_{0}|^{2}$, and $\theta$ is the angle
between the vector $\mathbf{k}$ of the primary UH wave and the
vector $\mathbf{k}-\mathbf{q}$ of the secondary UH wave,
\begin{equation}
\cos^{2}\theta=\frac{[\mathbf{k}\cdot
(\mathbf{k}-\mathbf{q})]^{2}}{k^{2}(\mathbf{k}-\mathbf{q})^{2}},
\,\, \sin^{2}\theta=\frac{[(\mathbf{k}\times
\mathbf{q})_{z}]^{2}}{k^{2}(\mathbf{k}-\mathbf{q})^{2}}.
\end{equation}
It can be seen that scalar and vector nonlinearities compete with
each other in parametric coupling. For example, at angles
satisfying $\tan\theta\gg (1+\mu^{2})/2\mu$, vector nonlinearity
dominates. Setting in Eq. (\ref{non-disp-decay1})
$\Omega=\Omega_{q}+\nu$, where $|\nu|\ll \Omega_{q}$, and assuming
that the resonance condition
$\omega_{\mathbf{k}}-\omega_{\mathbf{k}-\mathbf{q}}-\Omega_{q}=0$
is satisfied, we have
\begin{equation}
\label{non-disp-decay2}
\nu^{2}=-\frac{q^{2}|E_{0}|^{2}}{2\Omega_{q}\beta}
[(1+\mu^{2})^{2}\cos^{2}\theta+4\mu^{2}\sin^{2}\theta].
\end{equation}
The resonance condition corresponds to the decay of the UH wave
into another UH wave and an DIMA wave. From Eq.
(\ref{non-disp-decay2}) we obtain (by also transforming the
expression in square brackets) the decay instability growth rate,
\begin{equation}
\label{decay-growth2}
\gamma=q|E_{0}|\left\{\frac{[4\mu^{2}+(1-\mu^{2})^{2}\cos^{2}\theta]}
{2(\alpha+q^{2})\beta^{1/2}}\right\}^{1/2}.
\end{equation}
The maximum growth rate corresponds to angles $\theta$ satisfying
the condition $\cos^{2}\theta=1$. If $\Omega^{2}\gg\Omega_{q}^{2}$
and $\omega_{\mathbf{k}}-\omega_{\mathbf{k}-\mathbf{q}}=0$ we have
from Eq. (\ref{non-disp-decay1}),
\begin{equation}
\label{non-disp-decay3} \Omega^{3}=-\frac{q^{2}|E_{0}|^{2}}{\beta}
[4\mu^{2}+(1-\mu^{2})^{2}\cos^{2}\theta],
\end{equation}
and the growth rate is given by
\begin{equation}
\label{decay-growth3}
\gamma=\frac{\sqrt{3}\,q^{2/3}|E_{0}|^{2/3}}{2}
\left\{\frac{[4\mu^{2}+(1-\mu^{2})^{2}\cos^{2}\theta]}
{\beta}\right\}^{1/3}.
\end{equation}
This type of instability corresponds to the so-called modified
decay instability \cite{Zakharov1972}.

\subsection{Modulational instability}
At sufficiently large amplitudes of the plane pump wave, both
terms in the brackets of Eq. (\ref{non-disp}) are important, i.e.
both satellites are excited. In this case, instability (modulation
instability), in contrast to the previously considered case of
decay instability, has a threshold with respect to the pump
amplitude. In the most interesting case $q\gg k$, i.e. when the
wave numbers of perturbations are much greater than the wave
numbers of the plane pump wave, from Eqs. (\ref{D-pm}),
(\ref{non-disp}) and (\ref{C-pm}) we find,
\begin{gather}
(\Omega^{2}-\Omega^{2}_{q})(\Omega^{2}-q^{4})=\frac{2q^{4}|E_{0}|^{2}}{\beta}
\nonumber \\
\times[(1+\mu^{2})^{2}\cos^{2}\phi+4\mu^{2}\sin^{2}\phi],
\label{non-disp1}
\end{gather}
where $\phi$ is the angle between the vectors $\mathbf{k}$ and
$\mathbf{q}$,
\begin{equation}
\cos^{2}\phi=\frac{(\mathbf{k}\cdot \mathbf{q})^{2}}{k^{2}q^{2}},
\,\, \sin^{2}\phi=\frac{[(\mathbf{k}\times
\mathbf{q})_{z}]^{2}}{k^{2}q^{2}}.
\end{equation}
From Eq. (\ref{non-disp1}) it then follows,
\begin{gather}
\Omega^{2}=\frac{(\Omega_{q}^{2}+q^{4})}{2}
\nonumber \\
\pm
\left\{\frac{(\Omega_{q}^{2}-q^{4})^{2}}{4}+\frac{2q^{4}|E_{0}|^{2}
[4\mu^{2}+(1-\mu^{2})^{2}\cos^{2}\phi]}{\beta}\right\}^{1/2}.
\label{OM2}
\end{gather}
Instability occurs ($\Omega^{2}<0$) only when there is the sign
"$-$" before the curly brackets, and the pump amplitude exceeds
the threshold,
\begin{equation}
\label{threshold-main}
|E_{0}|^{2}>\frac{\alpha+q^{2}}{2[4\mu^{2}+(1-\mu^{2})^{2}\cos^{2}\phi]}.
\end{equation}
The first and second terms in the square brackets of the
denominator of Eq. (\ref{threshold-main}) correspond to vector and
scalar nonlinearity, respectively. The minimum instability
threshold
\begin{equation}
\label{threshold-main-min}
|E_{0}|_{\mathrm{th,min}}^{2}=\frac{\alpha+q^{2}}{2(1+\mu^{2})^{2}}
\end{equation}
corresponds to the wave vectors of perturbations $\mathbf{q}$
collinear with the pump wave vector $\mathbf{k}$. In contrast to
the usual modulation instability with local nonlinearity, the
instability thresholds are nonzero even for $q=0$. The instability
is purely growing (absolute instability) with the growth rate
$\gamma=\mathrm{Im}\,\Omega$, where $\Omega$ is determined in Eq.
(\ref{OM2}). This instability is an instability of a uniform field
(in the limit $\mathbf{k}\rightarrow 0$) leading to the splitting
of this field into clumps, which ultimately results in the
emergence of coherent structures at the nonlinear stage, which
generally speaking can be both non-stationary (collapsing
cavitons) and stationary (stable 2D solitons). As shown below, due
to nonlocal nonlinearity ($\alpha\neq 0$), it is precisely the
latter case that occurs. In the subsonic regime $\Omega^{2}\ll
\Omega_{q}^{2}$, the instability growth rate is
\begin{equation}
\label{growth-modul}
\gamma=q^{2}\left[\frac{2|E_{0}|^{2}[4\mu^{2}+(1-\mu^{2})^{2}\cos^{2}\phi]}{(\alpha+q^{2})}-1\right]^{1/2}.
\end{equation}
The maximum instability growth rate Eq. (\ref{growth-modul}) is
achieved at
\begin{equation}
q_{\mathrm{opt}}=\frac{1}{2}\left\{c-4\alpha+\left[c(c+8\alpha)\right]^{1/2}\right\}^{1/2},
\end{equation}
where
\begin{equation}
c=2|E_{0}|^{2}[4\mu^{2}+(1-\mu^{2})^{2}\cos^{2}\phi].
\end{equation}
At the nonlinear stage of instability, it is just on scales $\sim
1/q_{\mathrm{opt}}$ that one can expect the emergence of coherent
structures. In the case of long-wave modulations $q\ll k$, using
\begin{equation}
\omega_{\mathbf{k}\pm\mathbf{q}}\sim \omega_{\mathbf{k}}\pm
\frac{\partial
\omega_{\mathbf{k}}}{\partial\mathbf{k}}\cdot\mathbf{q}+\frac{1}{2}\frac{\partial^{2}
\omega_{\mathbf{k}}}{\partial\mathbf{k}^{2}}q^{2}
\end{equation}
in Eq. (\ref{D-pm}), from Eq. (\ref{non-disp}) we obtain
\begin{equation}
\label{convective1}
(\Omega^{2}-\Omega^{2}_{q})[(\Omega-\mathbf{q}\cdot\mathbf{v}_{g})^{2}-q^{4}]
=\frac{2(1+\mu^{2})^{2}q^{4}|E_{0}|^{2}}{\beta},
\end{equation}
where
$\mathbf{v}_{g}=\partial\omega_{k}/\partial\mathbf{k}=2\mathbf{k}$
is the group velocity of the UH wave. The contribution of vector
nonlinearity is absent in this case. Equation (\ref{convective1})
can be simplified in a number of cases. For example, if
$\Omega\ll\Omega_{q}$ we have from Eq. (\ref{convective1}),
\begin{equation}
(\Omega-\mathbf{q}\cdot\mathbf{v}_{g})^{2}
=q^{4}\left[1-\frac{2|E_{0}|^{2}(1+\mu^{2})^{2}}{\alpha+q^{2}}\right],
\end{equation}
and when the threshold
$|E_{0}|^{2}>(\alpha+q^{2})/2(1+\mu^{2})^{2}$ is exceeded,
instability occurs with the growth rate given by
\begin{equation}
\gamma=q^{2}\left[\frac{2|E_{0}|^{2}(1+\mu^{2})^{2}}{\alpha+q^{2}}-1\right]^{1/2}.
\end{equation}
Unlike the case of short-wave modulations of Eq. (\ref{OM2}), the
instability is not purely growing, but a convective instability
when growing disturbances are carried away with the group velocity
$\mathbf{v}_{g}$.

Numerical estimates for the threshold electric field in Eq.
(\ref{threshold-main-min}) for typical laboratory ($n_{e0}\sim
10^{9}$ cm$^{-3}$, $n_{d0}\sim 5\cdot 10^{5}$, $Z_{d}\sim 10^{4}$,
$B_{0}\sim 500$ G) and Martian ($n_{e0}\sim 10^{-3}$ cm$^{-3}$,
$Z_{d}n_{d0}\sim 0.9\cdot 10^{-2}$ cm$^{-3}$, $T_{e}\sim 5$ eV,
$B_{0}\sim 30$  $\mu$G) plasmas give $E\sim 10$ V/m and $E\sim 10$
$\mu$V/m, respectively.

\section{\label{soliton} Soliton solution}

Neglecting the time derivative on the left-hand side of Eq.
(\ref{mainLFdim}), that is, in the static (subsonic) regime, and
also neglecting the vector nonlinearity, which is valid if $\mu\ll
1$, Eqs. (\ref{mainHFdim}) and (\ref{mainLFdim}) become
\begin{gather}
\label{mainHF-r} \Delta\left(i\frac{\partial\varphi}{\partial t
}+\Delta\varphi\right)
 =\nabla\cdot(b\nabla\varphi),
\\
\label{mainLF-r} \alpha b-\Delta b=\Delta |\nabla\varphi|^{2}.
 \end{gather}
The system of Eqs. (\ref{mainHF-r}) and (\ref{mainLF-r}) can be
written as a single equation for the envelope potential $\varphi$,
\begin{equation}
\label{nonlocal-eq} \Delta\left(i\frac{\partial\varphi}{\partial t
}+\Delta\varphi\right)
 =\nabla\cdot\left\{\nabla\varphi
 \! \int \! \Delta G(\mathbf{r}-\mathbf{r}^{\prime})|\nabla\varphi (\mathbf{r}^{\prime})|^{2}
 d^{2}\mathbf{r}^{\prime}\right\},
 \end{equation}
where the kernel
 \begin{equation}
\label{R-kernel} G(\mathbf{r})=-
\frac{K_{0}(\sqrt{\alpha}|\mathbf{r}|)}{2\pi}
 \end{equation}
is the Green function of the 2D Helmholtz equation,
\begin{equation}
\label{Helmholtz}
(\alpha-\Delta)G(\mathbf{r})=-\delta(\mathbf{r}),
 \end{equation}
and $K_{0}(z)$ is the modified Bessel function of the second kind
of order zero. From Eq. (\ref{nonlocal-eq}) it can be clearly seen
that the nonlinearity is essentially nonlocal. The nonlocality
arises from the first term in Eq. (\ref{mainLF-r}) with
$\alpha\neq 0$, and in the physical sense is due to the gap in
dispersion of the DIMA wave, that is, the presence of the Rao
cutoff frequency. If $\alpha=0$ we have $\Delta
G(\mathbf{r})=\delta (\mathbf{r})$ and then Eq.
(\ref{nonlocal-eq}) is reduced to an equation with local
nonlinearity and it is completely analogous in form to the
Zakharov equation for nonlinear Langmuir waves
\cite{Zakharov1972}. The great diversity of dusty plasma
parameters under specific conditions (see, e.g,
\cite{Shukla-Mamun-book2002,Fortov2005}) leads to the fact that,
generally speaking, the nonlocality parameter $\alpha$ can be both
much less and much greater than unity (the first case is more
common). For example, for negatively charged dust particles we
have $n_{i0}/n_{e0}\gg 1$, and if $\mu\ll 1$ and
$v_{s}^{2}/v_{A}^{2}\sim 1$ the nonlocality parameter can be
estimated as $\alpha\sim
(\Omega_{i}^{2}/\omega_{pi}^{2})(n_{i0}/n_{e0})$ so that
$\alpha\gg 1$ for sufficiently strong magnetic fields and a strong
electron depletion due to a high negative charge concentration in
the plasma.

Equation (\ref{nonlocal-eq}) conserves the energy,
\begin{equation}
\label{norm2D} N=\int |\nabla\varphi|^{2}\,d^{2}\mathbf{r},
 \end{equation}
and Hamiltonian
 \begin{equation}
\label{hamiltonian} H=\!\!\int \!
\left\{|\Delta\varphi|^{2}-\frac{|\nabla\varphi|^{2}}{2}\! \int \!
\Delta G(\mathbf{r}-\mathbf{r}^{\prime})|\nabla\varphi
(\mathbf{r}^{\prime})|^{2}
 d^{2}\mathbf{r}^{\prime}\right\}d^{2}\mathbf{r},
 \end{equation}
and can be written in the hamiltonian form
\begin{equation}
i\frac{\partial}{\partial t}\Delta\varphi=\frac{\delta
H}{\delta\varphi^{\ast}}.
\end{equation}
For stationary solutions of the form
\begin{equation}
\varphi (\mathbf{r},t)=\Phi (\mathbf{r})\exp (i\lambda^{2}t),
\end{equation}
from Eqs. (\ref{mainHF-r}) and (\ref{mainLF-r}) we have
\begin{gather}
\label{mainHF-s} \Delta\left(-\lambda^{2}\Phi+\Delta\Phi\right)
 =\nabla\cdot(b\nabla\Phi),
\\
\label{mainLF-s} \alpha b-\Delta b=\Delta |\nabla\Phi|^{2}.
 \end{gather}
 \begin{figure}
\includegraphics[width=3.4in]{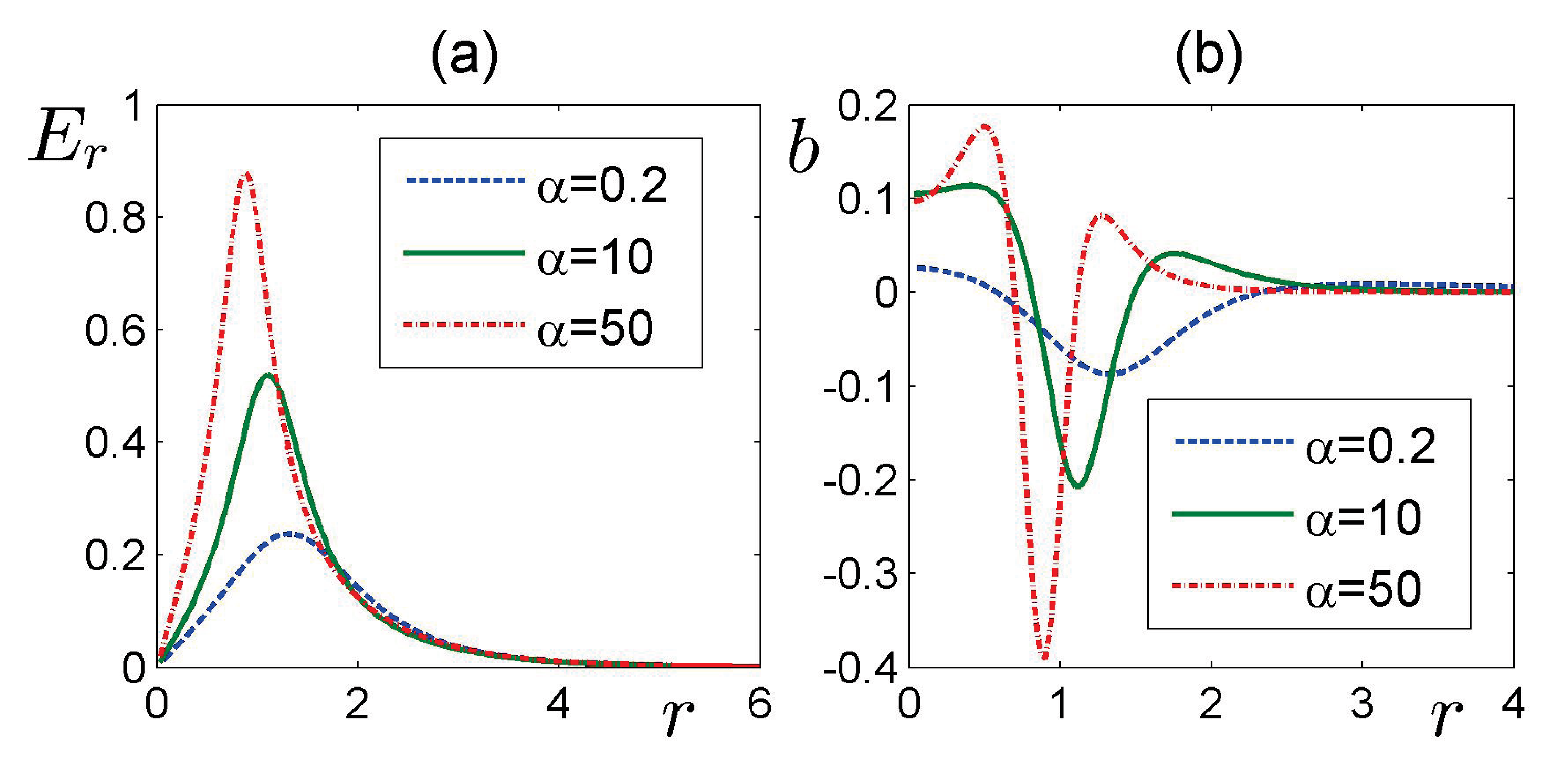}
\caption{\label{fig1} (a) Radial profiles of the radial electric
field $E_{r}$ and (b) the magnetic field perturbation $b$ for
different values of the nonlocality parameter $\alpha$. }
\end{figure}
 In the radially symmetric case, taking into account for the 2D radially symmetric
 Laplacian $\Delta=r^{-1}\partial_{r} (r\partial_{r})$, Eqs. (\ref{mainHF-s}) and
(\ref{mainLF-s}) can be reduced to
\begin{equation}
\label{HF-radial}
-\lambda^{2}E_{r}+\frac{d^{2}E_{r}}{dr^{2}}+\frac{1}{r}\frac{dE_{r}}{dr}-\frac{E_{r}}{r^{2}}=bE_{r}
\end{equation}
and
\begin{equation}
\label{LF-radial} \alpha
b-\left(\frac{d^{2}}{dr^{2}}+\frac{1}{r}\frac{d}{dr}\right)b
=\left(\frac{d^{2}}{dr^{2}}+\frac{1}{r}\frac{d}{dr}\right)
E_{r}^{2},
\end{equation}
respectively, where $E_{r}=\partial\Phi/\partial r$ is the radial
electric field. These equations are supplemented by boundary
conditions at zero and at infinity,
\begin{gather}
\label{boundary-zero} E_{r}=0, \quad \frac{db}{dr}=0, \quad
 \quad r=0,
\\
\label{boundary-infty} E_{r}\rightarrow 0, \quad b \rightarrow 0,
\quad \quad
 \quad \, r\rightarrow\infty .
 \end{gather}
 Equations (\ref{HF-radial}) and (\ref{LF-radial}) are solved numerically
 by the Petviashvili method
 \cite{Petviashvili_book1992,Pelinovsky2004,Lakoba2007},
 which has been successfully applied to find soliton solutions
 of autonomous nonlinear equations using the Fourier transform,
 and which is generalized (see Appendix A) for the case under consideration
 of explicit dependence on the spatial radial variable. Radial profiles of the radial electric
field $E_{r}$ and the magnetic field perturbation $b$ for
different values of the nonlocality parameter $\alpha$ are shown
in Fig.~1. It can be seen that, in contrast to the case of local
nonlinearity ($\alpha=0$), along with the magnetic well (and the
corresponding density well due to the frozen-in  field relation
Eq. (\ref{frozen})), there are also humps of magnetic field
perturbation (and density humps), that is, regions of space where
$b>0$. The presence of density humps together with a density well
is strikingly different, for example, from the known interaction
of HF Langmuir waves with LF ion-sound waves, when the nonlinear
effect is associated with a plasma density perturbation well.

\section{\label{stability} Ground state stability}

In this section we rigorously prove the stability of the 2D dust
soliton found in Sec. \ref{soliton} with respect to finite
perturbations, i.e. the Lyapunov stability. The essence of the
Lyapunov criterion for the stability of soliton structures
consists in the existence of a lower bound for the Hamiltonian
under the condition that other integrals of motion are fixed. We
follow the method of functional inequalities widely used to study
the stability of nonlinear stationary states
\cite{Rubenchik1986,Zakharov2012} and first applied by Zakharov
and Kuznetsov in Ref.~\cite{Zakharov1974} to prove the stability
of the three-dimensional ion-sound soliton in a magnetized plasma.

We represent the Hamiltonian (\ref{hamiltonian} ) as
\begin{equation}
\label{HHH} H=H_{0}-\frac{1}{2}H_{1},
\end{equation}
where
\begin{gather}
\label{H0} H_{0}=\int |\Delta\varphi|^{2}\,d^{2}\mathbf{r},
\\
\label{H1} H_{1}=\int \Delta
G(\mathbf{r}-\mathbf{r}^{\prime})|\nabla\varphi
(\mathbf{r})|^{2}|\nabla\varphi
(\mathbf{r}^{\prime})|^{2}\,d^{2}\mathbf{r}^{\prime}d^{2}\mathbf{r}.
\end{gather}
We rewrite $H_{1}$ as,
\begin{equation}
\label{H11} H_{1}=\int \Delta
G(\mathbf{r}-\mathbf{r}^{\prime})|\mathbf{r}-\mathbf{r}^{\prime}||\nabla\varphi
(\mathbf{r})|^{2}\frac{|\nabla\varphi
(\mathbf{r}^{\prime})|^{2}}{|\mathbf{r}-\mathbf{r}^{\prime}|}\,d^{2}\mathbf{r}^{\prime}d^{2}\mathbf{r}.
\end{equation}
Taking into account that $\Delta K_{0}(\alpha
|\mathbf{r}|)=K_{0}(\alpha |\mathbf{r}|)$ and introducing the
constant $C=\max_{z} [zG(z)]>0$ (since $G>0$), we have from Eq.
(\ref{H11}) an obvious inequality,
\begin{equation}
\label{H111} H_{1}\leq C\int |\nabla\varphi
(\mathbf{r})|^{2}\frac{|\nabla\varphi
(\mathbf{r}^{\prime})|^{2}}{|\mathbf{r}-\mathbf{r}^{\prime}|}\,d^{2}\mathbf{r}^{\prime}d^{2}\mathbf{r}.
\end{equation}
In turn, using H\"{o}lder inequality we have from Eq.
(\ref{H111}),
\begin{equation}
\label{H1111} H_{1}\leq C\int |\nabla\varphi
(\mathbf{r})|^{2}d^{2}\mathbf{r}\int\frac{|\nabla\varphi
(\mathbf{r}^{\prime})|^{2}}{|\mathbf{r}-\mathbf{r}^{\prime}|}\,d^{2}\mathbf{r}^{\prime}.
\end{equation}
Next, we make use of the inequality that represents one of the
versions of the Gagliardo-Nirenberg-Ladyzhenskaya inequality (the
proof is given in Ref.~\cite{Turitsyn1985}),
\begin{equation}
\label{aux-inequal} \int
\frac{f^{2}(\mathbf{r})}{|\mathbf{r}-\mathbf{r}^{\prime}|}d\mathbf{r}\leq
2\left(\int f^{2} d^{2}\mathbf{r}\right)^{1/2} \left(\int (\nabla
f)^{2} d^{2}\mathbf{r}\right)^{1/2},
\end{equation}
which is valid for an arbitrary sufficiently smooth function
$f(\mathbf{r})$. From Eq. (\ref{H1111}), using Eq.
(\ref{aux-inequal}) one can obtain,
\begin{gather}
H_{1}\leq 2C\left(\int |\nabla \varphi
(\mathbf{r})|^{2}d^{2}\mathbf{r}\right)^{3/2}\left(\int
|\Delta\varphi (\mathbf{r})|^{2}d^{2}\mathbf{r}\right)^{1/2}
\nonumber \\
 =2CN^{3/2}H_{0}^{1/2}.
 \label{H11111}
\end{gather}
Substituting Eq. (\ref{H11111}) into Eq. (\ref{HHH}) we arrive at
the following estimate for the Hamiltonian,
\begin{equation}
\label{H-bound} H\geq H_{0}-CN^{3/2}H_{0}^{1/2}
\end{equation}
Note that essentially the inequality (\ref{H-bound}) was obtained
in a different way for a similar model in Ref.~\cite{Sulem2009}.
Under the fixed energy $N$, the right-hand side of the inequality
(\ref{H-bound}) as a function of $H_{0}$ reaches its minimum at
$H_{0}=C^{2}N^{3}/4$, so that
\begin{equation}
\label{H-bound1} H\geq -\frac{C^{2}N^{3}}{4}.
\end{equation}
Thus, we have shown that, under the fixed conserved quantity $N$,
the Hamiltonian is bounded from below. From Eqs. (\ref{H-bound})
and (\ref{H-bound1}) it also follows that $H_{0}$ remains
uniformly bounded in time. In accordance with Lyapunov's theory,
due to the boundedness of the Hamiltonian from below, the
corresponding minimum is achieved at some stable configuration
corresponding to the ground state (2D soliton).

\section{Conclusion}

We have obtained a 2D nonlinear system of equations for the
electrostatic potential envelope and the LF magnetic field
perturbation to describe the interaction of the UH wave
propagating perpendicular to an external magnetic field with the
DIMA wave in a magnetized dusty plasma. The main nonlinear effect
is the action of the ponderomotive force of the HF pressure of the
UH wave on the LF motion of plasma, which in the linear case
correspond to the DIMA wave. This is reminiscent of the
interaction of HF Langmuir waves with LF ion-acoustic waves in a
non-magnetized plasma \cite{Zakharov1972}, where the ponderomotive
force of HF Langmuir waves leads to the formation of a plasma
density well and, in the three-dimensional case, to the phenomenon
of Langmuir wave collapse. In our case, in addition to the LF
plasma density perturbation, the LF magnetic field perturbation is
also taken into account. In addition, unlike
Ref.~\cite{Zakharov1972}, our equations contain both scalar and
vector nonlinearities, and, generally speaking, the contributions
of these nonlinearities are of the same order. The vector
nonlinearity has the form of a Poisson bracket and identically
disappears in the 1D case. Note also that, unlike
Ref.~\cite{Rao1995}, where the linear dispersion of the DIMA wave
was obtained in 1D form, we consider the 2D case for both the UH
and DIMA waves.

A nonlinear dispersion relation has been derived and decay and
modulation instabilities have been considered. The instability
growth rates and instability thresholds have been obtained in a
number of special cases. Numerical estimates of the modulation
instability thresholds for laboratory and Martian dusty plasmas
have been given, and these thresholds can easily be exceeded under
corresponding real physical conditions.

In the static (subsonic) approximation, a two-dimensional soliton
solution (ground state) has been found numerically by the
generalized Petviashvili method. The radial dependence of the
soliton profile for sufficiently large nonlocality parameters has
the form of a well with two humps. Such a peculiar form of the 2D
soliton has apparently not been obtained before. It should be
noted that in real physical situations involving dusty plasmas,
the nonlocality parameter can vary over very wide ranges.

We have shown that the presence of a gap in the DIMA wave
dispersion due to the Rao cutoff frequency causes the nonlinearity
to be nonlocal. The nonlocality of the nonlinearity is the key
point for the stability of the found two-dimensional soliton,
otherwise the soliton would either collapse or spread out. Using
the method of functional inequalities, we have shown that due to
nonlocal nonlinearity the Hamiltonian is bounded below at fixed
energy, thus proving the stability of the ground state against the
collapse.

We have restricted ourselves to the 2D case. It should be noted
that, as far as we know, the dispersion of the DIMA wave in the 3D
case, that is, when the wave propagates at an angle to the
external magnetic field, has not yet been considered. In this
regard, the question of the interaction of the UH wave and the
DIMA wave in the 3D case, when the condition of the almost
perpendicular propagation of the UH wave Eq. (\ref{cond-perp}) is
not valid, remains open.

\appendix

\section{Generalization of the Petviashvili method}
We follow Refs.~\cite{Lashkin2008Az2D,Lashkin2008Az3D}, where, in
particular, a generalization of the Petviashvili method for
finding stationary (soliton solutions) of nonlinear equations was
proposed for the case when the equations contains an explicit
dependence on spatial variables.

Let us consider a system of nonlinear equations for two fields $u$
and $b$ ($u$ can be complex),
\begin{gather}
\label{L1} L_{1}u=N_{1}[u,u^{\ast},b],
\\
\label{L2} L_{2}b=N_{2}[u,u^{\ast}],
\end{gather}
where $L_{1}$ and $L_{2}$ are linear operators, and $N_{1}$ and
$N_{2}$ account for the nonlinear terms. The spatial dimension $D$
is arbitrary. Generalization to a larger number of fields is quite
transparent and does not present any difficulties. Generally
speaking, the system of Eqs. (\ref{L1}) and (\ref{L2}) is not
assumed to be autonomous, that is, it may contain explicit
dependencies on spatial coordinates. A special case of Eqs.
(\ref{L1}) and (\ref{L2}) is the system of nonlinear equations
(\ref{HF-radial}) and (\ref{LF-radial}) considered in this paper.
At each iteration step $n$, the linear Eqs. (\ref{L1}) and
(\ref{L2}) are solved using the known $u$, $u^{\ast}$, and $b$ on
the right-hand sides of Eqs. (\ref{L1}) and (\ref{L2}). First,
using the known $u^{(n)}$, the Eq. (\ref{L2}) is solved to obtain
$b^{(n)}$, that is,
\begin{equation}
b^{(n)}=L_{2}^{-1}N_{2}[u^{(n)},u^{\ast, (n)}].
\end{equation}
Secondly, using the already known value $b^{(n)}$, Eq. (\ref{L1})
is solved to find a new approximation $\hat{u}^{(n)}$, that is,
\begin{equation}
\hat{u}^{(n)}=L_{1}^{-1}N_{1}[u^{(n)},u^{\ast, (n)},b^{(n)}].
\end{equation}
Then the iteration procedure at the $n$-th iteration is
\begin{equation}
u^{(n+1)}=s\hat{u}^{(n)},
\end{equation}
where $s$ is the so-called stabilizing factor defined by
\begin{equation}
s=\left(\frac{\int |u^{(n)}|^{2}d^{D}\mathbf{r}}{\int
|u^{(n)}\hat{u}^{(n)}|d^{D}\mathbf{r}}\right)^{\gamma},
\end{equation}
and $\gamma>1$. The progressive iterations are terminated when
$|s-1|<\epsilon$, and typically $\epsilon=10^{-13}-10^{-15}$. For
the power nonlinearity, the fastest convergence is achieved for
$\gamma=p/(p-1)$, where $p$ is the power of nonlinearity. In our
case of cubic nonlinearity we have $\gamma=3/2$. For nonlinearity
other than power-law, the value of $\gamma$ corresponding to the
fastest convergence is selected empirically, but in any case
$1<\gamma<p/(p-1)$, where $p$ is the smallest exponent in the
Taylor series expansion of nonlinearity. Note that the
Petviashvili iterative procedure always converges to the ground
state regardless of the initial guess, which may have a form far
from soliton and even have a different topology.

The linear equations $L_{1}u=f_{1}$ and $L_{2}b=f_{2}$, where
$f_{1}$ and $f_{2}$ are the corresponding right-hand sides, can be
solved in different ways. For example, after representing them
with the aid of a finite difference scheme, the solutions can be
found using iterative methods \cite{Saad1996} or direct matrix
solvers \cite{Duff1986}. Specifically, the discretization of
differential operators in Eqs. (\ref{HF-radial}) and
(\ref{LF-radial}) on a spatial grid with a grid spacing $h$ and an
accuracy $O(h^{2})$  has the form
\begin{equation}
\frac{d^{2}U}{dr^{2}}\rightarrow
\frac{U_{i+1}+U_{i-1}-2U_{i}}{h^{2}}, \quad
\frac{dU}{dr}\rightarrow \frac{U_{i+1}-U_{i-1}}{2h},
\end{equation}
where $U$ is the corresponding field, and $i$ is the grid node
number. Then the solution of the corresponding linear equations is
reduced to the inversion of tridiagonal matrices.

For autonomous Eqs. (\ref{L1}) and (\ref{L2}), using the Fourier
transform, we recover the conventional Petviashvili method.

\end{document}